\pdfoutput=1
\documentclass[twocolumn,superscriptaddress]{revtex4-1}

\usepackage{float}
\usepackage{amsmath}
\usepackage{amssymb}
\usepackage{amsfonts}
\usepackage{amsthm}
\usepackage{bm}
\usepackage{braket}
\usepackage{cases}
\usepackage[pdftex]{graphicx,color}
\usepackage{subfigure}

\newtheorem{theo}{Theorem}

\def\({\left(}
\def\){\right)}

\allowdisplaybreaks[2]

\begin{document}
\title{Rigorous calibration method for photon-number statistics}
\author{Masahiro Kumazawa}
\affiliation{Photon Science Center, Graduate School of Engineering, The University of Tokyo,
7-3-1 Bunkyo-ku, Tokyo 113-8656, Japan}
\affiliation{Department of Applied Physics, Graduate School of Engineering,
The University of Tokyo, 7-3-1 Hongo, Bunkyo-ku, Tokyo 113-8656, Japan}
\author{Toshihiko Sasaki}
\affiliation{Photon Science Center, Graduate School of Engineering, The University of Tokyo,
7-3-1 Bunkyo-ku, Tokyo 113-8656, Japan}
\author{Masato Koashi}
\affiliation{Photon Science Center, Graduate School of Engineering, The University of Tokyo,
7-3-1 Bunkyo-ku, Tokyo 113-8656, Japan}
\affiliation{Department of Applied Physics, Graduate School of Engineering,
The University of Tokyo, 7-3-1 Hongo, Bunkyo-ku, Tokyo 113-8656, Japan}

\begin{abstract}
 Characterization of photon statistics of a light source is one of the most basic tools in quantum optics.
 Although the outcome from existing methods is believed to be a good approximation when the measured light is sufficiently weak,
 there is no rigorous quantitative bounds on the degree of the approximation.
 As a result, they fail to fulfill the demand arising from emerging applications of quantum information such as quantum cryptography.
 Here, we propose a calibration method to produce rigorous bounds for a photon-number probability distribution
 by using a conventional Hanbury-Brown-Twiss setup with threshold photon detectors.
 We present a general framework to treat any number of detectors and non-uniformity of their efficiencies.
 The bounds are conveniently given as closed-form expressions of the observed coincidence rates and the detector efficiencies.
 We also show optimality of the bounds for light with a small mean photon number.
 As an application, we show that our calibration method can be used for the light source in a decoy-state quantum key distribution protocol.
 It replaces the {\it a priori} assumption on the distribution that has been commonly used,
 and achieves almost the same secure key rate when four detectors are used for the calibration.
\end{abstract}

\maketitle

\section{Introduction}
Autocorrelation measurement of a light source has been known to be a convenient method for investigating the characteristics of the source.
It dates back to Hanbury-Brown and Twiss (HBT) who measured \cite{HanburyBrown1956,HanburyBrown1957} the correlation of photocurrents from two photodetectors shone
by a common light source to determine its second-order intensity correlation function.
The meaning of the correlation functions in quantum optics was clarified by Glauber \cite{Glauber1963,Glauber1963a},
who showed that the second-order correlation function $G^{(2)}(r_1,r_2; \tau)=\braket{I(r_1,t)I(r_2,t+\tau)}$ can also be determined from a HBT setup with two photon detectors by measuring the rate of coincidence counts.
It was successfully used in the direct observation of the non-classical property of light \cite{Kimble1977}.
In the case of a pulsed light, a proper integration around $\tau=0$ and a normalization lead to a normalized factorial moment $g^{(2)}(0)=\braket{n(n-1)}/ \braket{n}^2$ of the photon number $n$ in the pulse \cite{Koashi1993}.
Since $g^{(2)}(0)=0$ is achieved only by an ideal single photon source (SPS), the measurement of $g^{(2)}(0)$ by the HBT setup has been widely adopted for the characterization of experimentally developed SPSs \cite{Lounis2005,Buckley2012}.
The extension of the method to higher-order moments $g^{(r)}(0)$ is also straightforward by increasing the number of detectors to $r$ \cite{Stevens2014,Rundquist2014}.

Despite the apparent success for the non-classical light sources, the above characterization method is not particularly suited to the emergent applications in quantum information.
The most severe drawback is the fact that the HBT setup with conventional threshold photon detectors provides the accurate value of the factorial moment only in the limit of low detection efficiencies. 
The value of $g^{(r)}(0)$ from an actual experiment is only approximate, and how it is deviated from the true value is unknown.
This is problematic for the applications involving the security\cite{Gottesman2004,Dunjko2012}, for which rigorous security statements are mandatory.
For the computational tasks using photons such as boson sampling \cite{Aaronson2013}, reliability of the outcome will eventually be ascribed to that of the constituent components including light sources.
Another drawback is that we encounter the moments $\{g^{(r)}(0)\}_r$ much less frequently in the quantum information theory than the photon-number probability distribution $\{p_n\}_n$.
Photon-based protocols concern the presence of a photon in a pulse, and hence the relevant quantity is the corresponding probability $p_1$.
The requirement for the light source used in the decoy-state quantum key distribution (QKD) \cite{Hwang2003,Wang2005,Lo2005} is given by a set of inequalities in terms of the photon-number probability distribution $\{p_n\}_n$ \cite{Adachi2007,Wang2009}.
While the knowledge on $\{g^{(r)}(0)\}_r$ may be converted to bounds on $\{p_n\}_n$, it is not straightforward due to the unboundness of the photon number $n$.
Hence, as a calibration method in the era of quantum information, it is vital that it provides rigorous statements over $\{p_n\}_n$,
and preferably it is tight.

In this paper, we propose a calibration method to obtain rigorous bounds on $\{p_n\}_{n\leq D-1}$ using a HBT setup with $D$ threshold photon detectors. It is flexible and versatile since the formula for an arbitrary number of detectors is given in a closed-form expression and the detection efficiencies do not need to be uniform. The explicit formula enables us to cope with various imperfections such as ambiguity in detection efficiencies and statistical uncertainty in the outcomes.
Our method is optimal for coherent states and thermal states with a small mean photon numbers.
As an application of our calibration methods to QKD, we modify the security analysis of the decoy-state BB84 protocol to accommodate the use of a calibrated source. It serves to bolster the security by lifting the {\it a priori} assumption on the source, and we further show that the performance drop is minimal when a $D=4$ setup is used for the calibration.

\section{Calibration method for the photon number distribution of a photon source}
\begin{figure}[t]
  \begin{center} 
    \includegraphics[width=8cm]{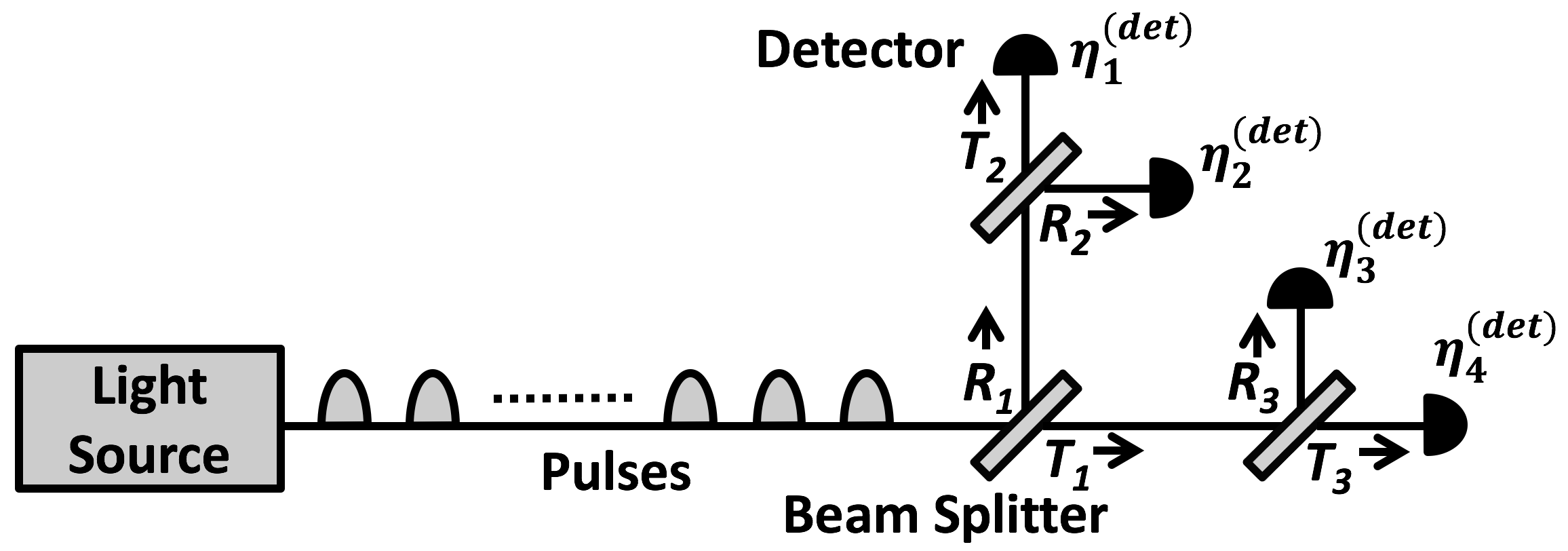}
	 \caption{
	 An implementation of the $D=4$ case of our calibration method.
	 The overall detection efficiencies are given by $\eta_1=R_1T_2\eta_1^{(\mathrm{det})}$, $\eta_2= R_1R_2\eta_2^{(\mathrm{det})}$, $\eta_3= T_1R_3\eta_3^{(\mathrm{det})}$, and $\eta_4=T_1T_3\eta_4^{(\mathrm{det})}$, where $T_{k}$ $(k=1, \ldots, 3)$ and $R_{k}$ $(k=1, \ldots, 3)$ are transmittance and reflectance, respectively.}
    \label{fig:detecter1} 
  \end{center}
\end{figure}

We consider a generalized HBT setup with $D$ photon detectors,  where the light pulses to be calibrated are divided into $D$ parts by beam splitters, and various rates of coincidence detections are recorded.
We assume that each detector is a threshold detector with quantum efficiency $\eta_i^{(\mathrm{det})}$,
which is modeled as reporting a presence of nonzero photons after a linear absorber with transmission $\eta_i^{(\mathrm{det})}$.
Figure\;\ref{fig:detecter1} shows an example for $D=4$.
The configuration of the beam splitters is not relevant.
The only relevant system parameters are the overall efficiency for each detector, $\eta_i\; (i=1,\ldots,D)$, including the branching efficiencies.
We denote their average as $\eta:=D^{-1}\sum_i \eta_i$.

We define the averaged $r$-tuple coincidence probability $c_{\mathrm{obs}, r}$ to represent the observed data. For the $r$-tuple coincidence, there are $\binom{D}{r}$ different combinations of $r$ detectors, and the associated probabilities. The averaged quantity $c_{\mathrm{obs}, r}$ is their arithmetic mean.
Let us first consider the case where the measured pulse contains exactly $n$ photons. The coincidence detection probability of the first to the $r$-th detectors is given by $\sum_{W\subset Z_r}(-1)^{|W|}(1-\sum_{i\in W}\eta_{{i}})^n$ where $Z_r:=\{1,2,\ldots,r\}$ and $|W|$ is the cardinality of set $W$.
The averaged probability $c_{\mathrm{obs}, r}$ for the $n$-photon input is then given by
\begin{align}
c_{n,r}:=\sum_{j=0}^{r}(-1)^{j}\omega_{r,j}\sum_{W\in{I}_j}(1-\sum_{i\in W}\eta_{i})^n \label{eq:cnr}
\end{align}
where we defined $\omega_{r,j}:=\binom{D-j}{r-j}/\binom{D}{r}$ and ${I}_j:=\{W\subset Z_D\ |\ |W|=j$\}.
Note that $c_{n,r}=0$ if $r>n$.
For the case of a general input pulse with distribution $\{p_n\}_n$, the averaged coincidences should satisfy
\begin{align}
   {c}_{{\rm obs},r}=\sum_{n=0}^{\infty}p_{n}{c}_{n,r} \ (r=1,\ldots,D). \label{eq:constraint}
\end{align}
Our goal is to find rigorous bounds on $\{p_n\}_n$ under the constraint of Eq.\;(\ref{eq:constraint}).

To describe the formula for the bounds, it is convenient to introduce vectors of order $D+1$ as follows.
Let $\bm{c}_{{\rm obs}}:=(1,c_{{\rm obs},1},\ldots,c_{{\rm obs},D})$, $\bm{c}_{n}:=(1,c_{n,1},\ldots,c_{n,D})$, and $\bm{c}_{\infty}:=(1,1,\ldots,1)$.
Consider a basis $\{\bm{c}_j\}_{j\in S}$ specified by the index set $S=\{0,1,\ldots,D\}$, and let $\{\bm{d}_j^{(S)}\}_{j\in S}$ be its reciprocal basis, namely, $\bm{c}_{j}\cdot\bm{d}^{(S)}_i=\delta_{ij}$ for $i,j\in S$.
Similarly define $\bm{d}^{(S')}_i$ for $S':=\{0,1, \ldots , D-1, \infty\}$. 
We can obtain $\{\bm{d}^{(S)}_i\}_i$ and $\{\bm{d}^{(S')}_i\}_i$ as the rows of inverses of $C:=(\bm{c}_{0}^{\mathrm{T}}, \bm{c}_{1}^{\mathrm{T}} ,\ldots, \bm{c}_{D}^{\mathrm{T}})$ and $C':=(\bm{c}_{0}^{\mathrm{T}}, \bm{c}_{1}^{\mathrm{T}} ,\ldots, \bm{c}_{D-1}^{\mathrm{T}}, \bm{c}_{\infty}^{\mathrm{T}})$, respectively. We postulate that the variations among the efficiencies $\{\eta_i\}_i$ are moderate, and they satisfy 
\begin{align}
\sum_{i\in W}\eta_i<\sum_{i\in W'}\eta_i\ \ \ {\rm if}\ \ |W|<|W'| \label{eq:eta}
\end{align}
for any $W, W' \subset Z_D$.
Now our main result is stated in the form of the following theorem.\\

\begin{theo}
 Under the condition (\ref{eq:eta}), the constraint (\ref{eq:constraint}) implies 
\begin{numcases}
   {p_n \leq p_n^U :=}
	 \bm{c}_{{\rm obs}}\cdot\bm{d}^{(S)}_n \ \ (D-n : {\rm even}) \label{eq:pueven} \\
	 \bm{c}_{{\rm obs}}\cdot\bm{d}^{(S')}_n \ \ (D-n : {\rm odd}) \label{eq:puodd}
 \end{numcases} 
 \begin{numcases}
   {p_n \geq p_n^L :=}
	 \bm{c}_{{\rm obs}}\cdot\bm{d}^{(S')}_n \ \ (D-n : {\rm even}) \label{eq:pleven} \\
	 \bm{c}_{{\rm obs}}\cdot\bm{d}^{(S)}_n \ \ (D-n : {\rm odd}) \label{eq:plodd}
 \end{numcases}
for $n=0,\ldots, D-1$ and
\begin{align}
\sum_{n=D}^{\infty} p_n \leq p^U_{\geq D} := \bm{c}_{{\rm obs}}\cdot\bm{d}^{(S)}_D. \label{eq:pud}
\end{align}
\begin{equation}
 \sum_{n=D}^{\infty} p_n \geq p^L_{\geq D} := \bm{c}_{{\rm obs}}\cdot\bm{d}^{(S')}_\infty . \label{eq:pld}
\end{equation}
\end{theo}

We first prove the inequalities (\ref{eq:pueven}), (\ref{eq:plodd}), and (\ref{eq:pud}) involving $S$. The crux of the proof is to show that $\bm{c}_{m}\cdot\bm{d}^{(S)}_{n}$ has a constant sign for $m>D$.
To do so, we focus on the property of $\bm{c}_{m}\cdot\bm{d}^{(S)}_{n}$ as a function of $m$ for a fixed value of $n$. Let us introduce a smooth function $f(m)$ over $\mathbb{R}$ as
\begin{align}
f(m):=z_{0}+\sum_{j=1}^{D}(-1)^{j}z_{j}\sum_{W\in{I}_j}e^{-\alpha_{W}m}, \label{eq:fm}
\end{align}
with $\alpha_{W}:=-{\rm ln}(1-\sum_{i\in W}\eta_{i})>0$.
Assume that the constants $\{z_j\}_j$ are given by $z_j = \bm{\omega}_j \cdot \bm{d}_n^{(S)}$, where $\bm{\omega}_j:= (\omega_{0,j},\omega_{1,j},\ldots, \omega_{D,j} )$ with $\omega_{r,j}:=0$ for $r<j$.
From Eq.\;(\ref{eq:cnr}), we see that $f(m)=\bm{c}_{m}\cdot\bm{d}^{(S)}_{n}$ for nonnegative integer $m$. By definition of $\bm{d}^{(S)}_{n}$, $f(n)=1$ and $f(m)=0$ for $m \in S\setminus\{n\}$.

Let us show that the numbers of zeros of $f(m)$ and its derivative $f'(m)$ are $D$ and $D-1$, respectively. The prerequisite (\ref{eq:eta}) ensures that there exists $\{\beta_j\}_{j=1}^{D-1}$ such that ${\rm max}_{W\subset{I}_j}{\alpha_{W}}<\beta_j<{\rm min}_{W\subset{I}_{j+1}}{\alpha_{W}}$. Let us temporarily assume that all $\{z_j\}_j$ are nonzero and have the same sign. Using the notation $D_{\gamma}:=e^{-\gamma m}(\frac{\mathrm{d}}{\mathrm{d}m})e^{\gamma m}$, we have
\begin{equation}
 D_{\beta_{1}}D_{0}f(m) = -\sum_{j=1}^{D}\sum_{W\in{I}_j}(-1)^j(\beta_1-\alpha_W)\alpha_W z_je^{-\alpha_{W}m}. \label{eq:DDf}
\end{equation}
We see that the coefficients of $e^{-\alpha_W m}$ for $j=1$ and $2$ have the same sign. In a similar vein, all the coefficients in ${D}_{\beta_{D-1}}\cdots {D}_{\beta_{1}}{D}_{{0}} f(m)$ have the same sign, implying that this function has no zeros. Since multiplication of $e^{\pm \gamma m}$ does not change zeros, Rolle's theorem assures that $f(m)$ should have no more than $D$ zeros. If some of $\{z_j\}_j$ were zero or had the opposite sign, we could remove the zeros by a fewer number of operating $D_\gamma$, contradicting the fact that $f(m)$ has at least $D$ zeros. Hence, we conclude that $f(m)$ has exactly $D$ zeros at $m \in S\setminus\{n\}$, and $f'(m)$ has exactly $D-1$ zeros.

Since Rolle's theorem also implies that the $D-1$ zeros of $f'(m)$ must be strictly between neighboring zeros of $f(m)$, it follows that $f'(m)\neq 0$ for $m \in S\setminus\{n\}$. Hence, $f(m)$ changes its sign across every point in $S\setminus\{n\}$. When $D-n$ is even, $f(n)=1$ then implies $f(m)>0$ for $m>D$.
Then we have $\bm{c}_{\rm obs}\cdot\bm{d}^{(S)}_{n}= \sum_m p_m f(m) \geq p_n$, proving Eq.\;(\ref{eq:pueven}). The case with $D-n$ being odd similarly leads to Eq.\;(\ref{eq:plodd}).

For the special case of $n=D$, the largest zero of $f'(m)$ lies in $(D-2, D-1)$. Since $f(D-1)=0$ and 
$f(D)=1$, we have $f'(m) >0$ for $m\geq D-1$, and hence $f(m)\geq 1$ for $m\geq D$. This leads to Eq.\;(\ref{eq:pud}).

The remaining inequalities (\ref{eq:puodd}),(\ref{eq:pleven}) and (\ref{eq:pld}) involving the index set $S'$ can be proved in a similar way.
For Eqs.\;(\ref{eq:puodd}) and (\ref{eq:pleven}), the function $f(m)$ has only $D-1$ known zeros at $S'\setminus\{n,\infty \}$, but it is compensated from the fact that $z_0=0$. The detail is given in Appendix \ref{appendix-proof-theorem}, which concludes the proof of Theorem 1.

\begin{figure*}[t]
 \begin{minipage}{0.98\columnwidth}
  \begin{center}
   \subfigure[$p_0$]{\includegraphics[width=70mm]{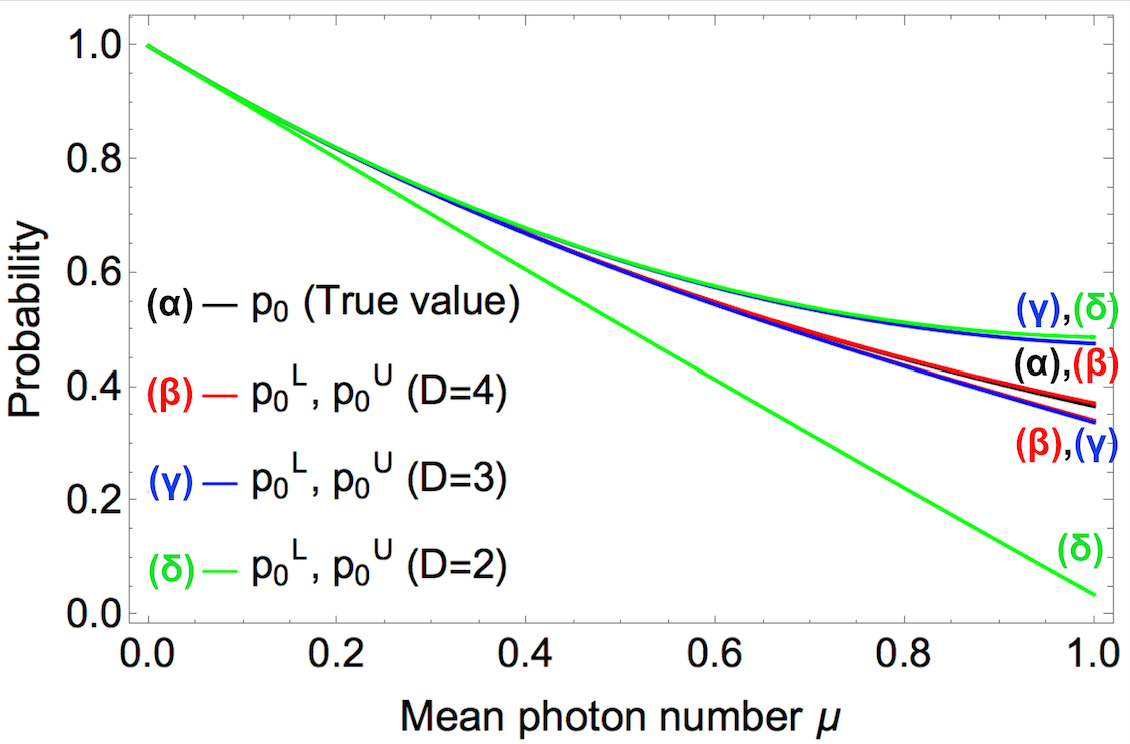}}
  \end{center}
 \end{minipage}
 \begin{minipage}{0.98\columnwidth}
  \begin{center}
   \subfigure[$p_1$]{\includegraphics[width=70mm]{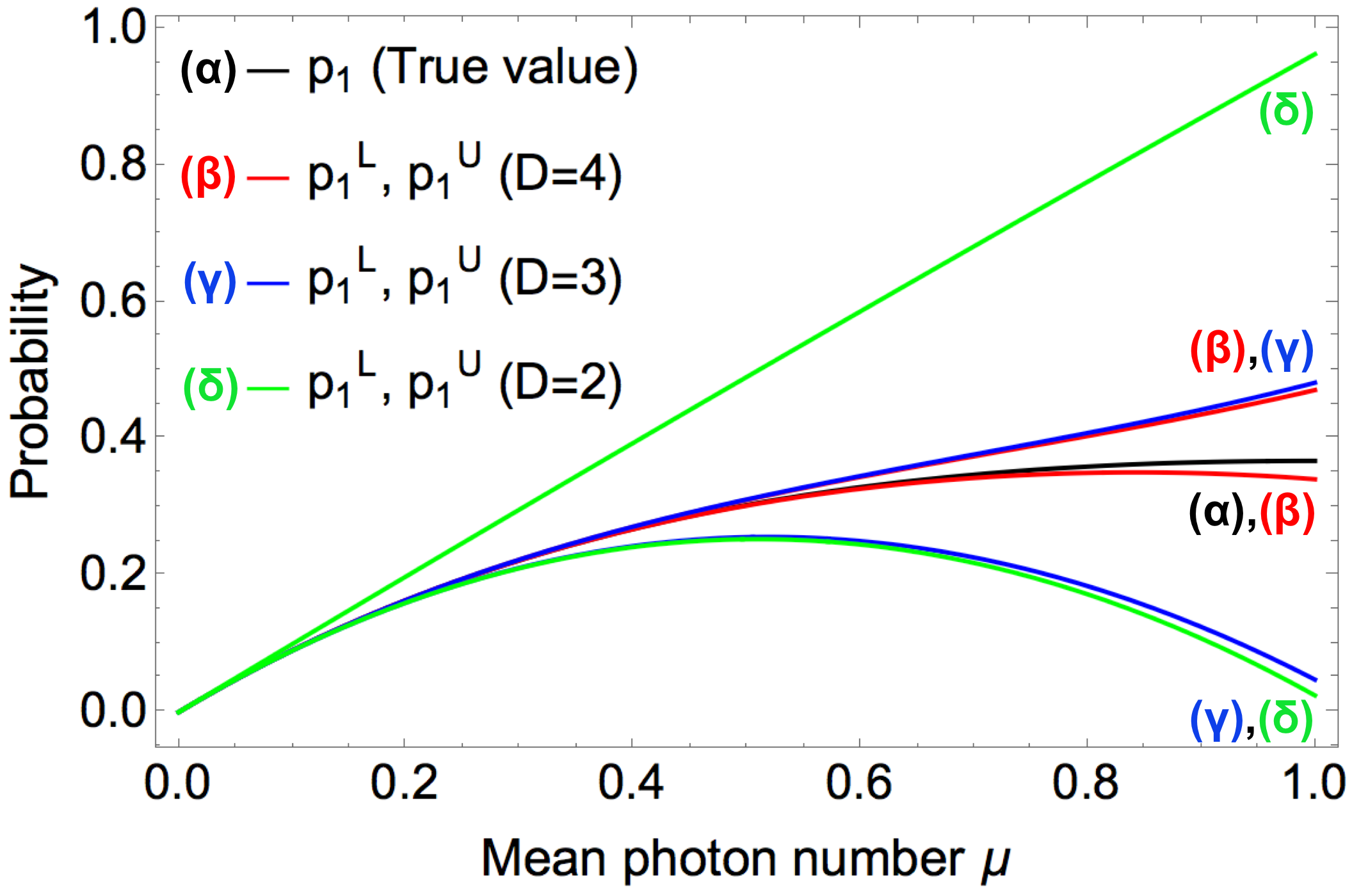}}
  \end{center}
 \end{minipage}

 \begin{minipage}{0.98\columnwidth}
  \begin{center}
   \subfigure[$p_2$]{\includegraphics[width=70mm]{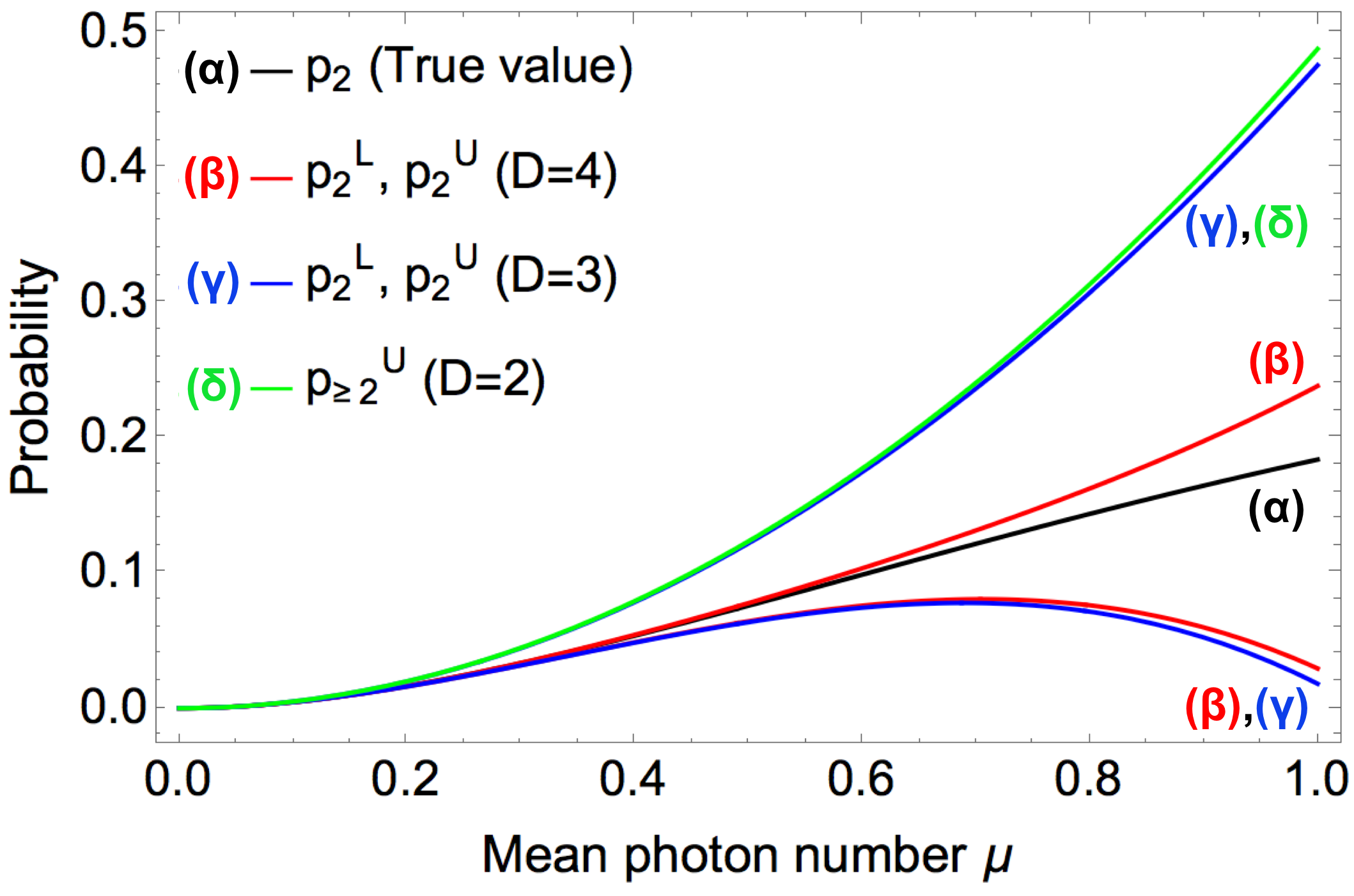}}
  \end{center}
 \end{minipage}
 \begin{minipage}{0.98\columnwidth}
  \begin{center}
   \subfigure[$p_3$]{\includegraphics[width=70mm]{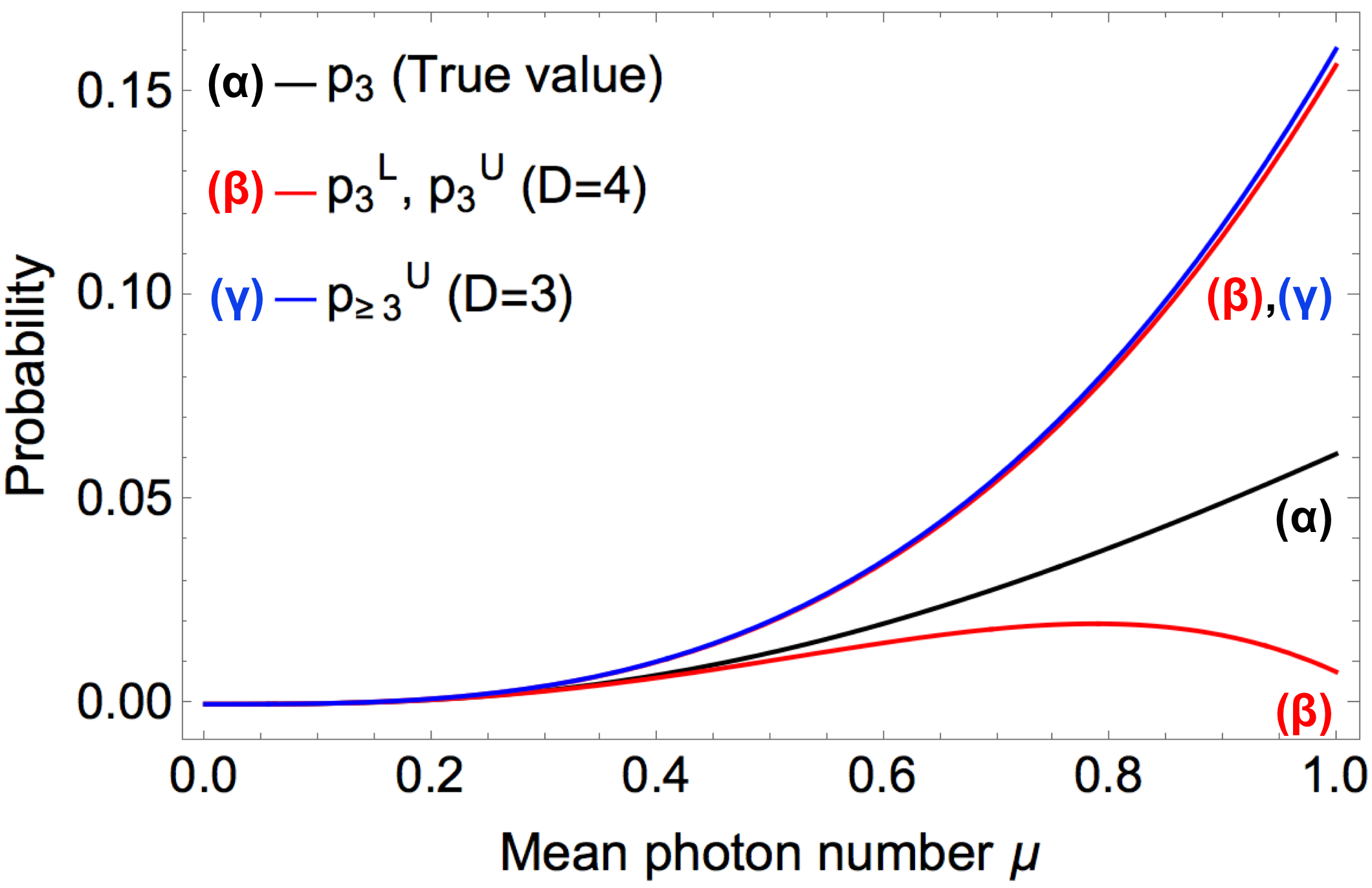}}
  \end{center}
 \end{minipage}
  \caption{(a)-(d) The bounds from the calibration method with $D=4$ and $\eta_1=\eta_2=\eta_3=\eta_4=0.025$ when applied to an ideal Poissonian source with $p_{n}=e^{-\mu}\mu^n/n!$ where $\mu$ is the mean photon number.}
    \label{fig:p0123}
\end{figure*}

Calculation of the bounds $p^L_n$ and $p^U_n$ in Eqs.\;(\ref{eq:pueven})-(\ref{eq:pld}) is straightforward 
because $C$ and $C'$ are triangular matrices.
Closed-form expressions for $D=2, 3$ and $4$ are given in Appendix \ref{appendix-explicit-formula}
in terms of normalized quantities $\tilde{c}_{\mathrm{obs},r}:= c_{\mathrm{obs},r}/c_{r,r}$ for convenience,
since they are $O(1)$ in the limit of $\eta_i\to 0$.
In Fig.\;\ref{fig:p0123}, we show the bounds obtained from the $D=4$ setup when it is applied to
an ideal Poissonian source with $p_{n}=e^{-\mu}\mu^n/n!$ along with the true values.
As shown in Fig.\;\ref{fig:p0123}, the bounds are fairly good for $\mu \leq 0.5$, and gradually become worse as $\mu$ gets larger.

When we compare the explicit formulas for different values of $D$, we notice that sometimes the dominant $O(1)$ terms in $\eta$ do not change as an increment of $D$, like Eq.\;(B2) to Eq.\;(B8), and Eq.\;(B10) to Eq.\;(B18).
In general, we can prove that ${\bm c}_{\text{obs}}\cdot {\bm d}_n^{(S)}$ for $D=D_0$ and ${\bm c}_{\text{obs}}\cdot {\bm d}_n^{(S')}$ for $D=D_0+1$ coincide in the limit of $\eta\to 0$ for a given source (see Appendix \ref{appendix-relation-eta0}).
In other words, each bound $p_n^{U,L}$ shows a major improvement for every other increment of the number $D$ of detectors used in the set up.
The origin of this unexpected behavior may be understood from the threshold or saturation behavior of the detectors, which an adversary may exploit to fool us to believe in a wrong distribution.
Given a distribution $\{p_n\}_n$ and the corresponding $\{c_{\text{obs},r}\}_r$, one may change $\{p_n\}_n$ only slightly by replacing a $O(\eta^D)$ potion of the pulses with ones with an extremely large photon number to flood the $D$ detectors.
Through this small modification, the adversary can increase $c_{\text{obs}, D}$ to any value without changing the rest of $\{c_{\text{obs},r}\}_r$.
Existence of such an attack forces us to trust the observed value of $c_{\text{obs}, D}$ only in one direction, which results in improving only half of the bounds.
The above argument tells us that, in the derivation of the rigorous bound, it is essential to take the saturation behavior of the detectors into consideration.

The present formula allows us to check the optimality of the obtained bounds. From the first row of the equality $C C^{-1}\bm{c}^{\mathrm{T}}_{\mathrm{obs}}=\bm{c}^{\mathrm{T}}_{\mathrm{obs}}$, we have $\sum^D_{n=0} \bm{c}_{\mathrm{obs}}\cdot \bm{d}^{(S)}_{n} = 1$. Hence, if $0\leq \bm{c}_{\mathrm{obs}}\cdot \bm{d}^{(S)}_{n}$ for $0\leq n \leq D$ (or equivalently, if all the $S$-related bounds are nonnegative), we find that $\{p_n\}_n$ defined as $p_n = \bm{c}_{\mathrm{obs}}\cdot \bm{d}^{(S)}_{n}$ for $0\leq n\leq D$ and $p_n=0$ for $n\geq D+1$ is a probability distribution, and it fulfills Eq.\;(\ref{eq:constraint}) as is seen from the remaining rows of $C C^{-1}\bm{c}^{\mathrm{T}}_{\mathrm{obs}}=\bm{c}^{\mathrm{T}}_{\mathrm{obs}}$. The inequalities (\ref{eq:pueven}),(\ref{eq:plodd}) and (\ref{eq:pud}) can thus be simultaneously saturated and no tighter bounds exist.
Similarly, if $\bm{c}_{\mathrm{obs}}\cdot\bm{d}^{(S')}_{n}\geq 0$ for $0\leq n\leq D-1$ and $\bm{c}_{\mathrm{obs}}\cdot\bm{d}^{(S')}_{\infty}\geq 0$, we can construct a probability distribution defined by $p_n = \bm{c}_{\mathrm{obs}}\cdot \bm{d}^{(S')}_{n}$ for $0\leq n\leq D-1$, $p_n= \bm{c}_{\mathrm{obs}}\cdot\bm{d}^{(S')}_{\infty}$ for $n=n_0$, and $p_n=0$  otherwise.
It saturates the inequalities (\ref{eq:puodd}),(\ref{eq:pleven}) and (\ref{eq:pld}), while it fulfills Eq.\;(\ref{eq:constraint}) in the limit of $n_0\to \infty$, implying the optimality of Eqs.\;(\ref{eq:puodd}),(\ref{eq:pleven}) and (\ref{eq:pld}).

Thanks to the closed-form expression of the bounds, we may discuss what types of light sources are optimally calibrated.
We consider the limit of $\eta\to 0$.
For weak light sources where $p_n$ rapidly decreases with $n$, the most severe condition for the optimality of the $S$-related bounds is expected to be
$p_{D-1}^L=\bm{c}_{\text{obs}}\cdot\bm{d}^{(S)}_{D-1} \geq 0$.
Indeed, if a light source satisfies
\begin{equation}
 \label{optimal-cond-physical-add}
	p_n > \frac{1}{D}\binom{D}{n}p_{D-1}
\end{equation}
for $n=D-3,D-5,\ldots$,
we can show that the condition $\bm{c}_{\text{obs}}\cdot\bm{d}^{(S)}_{D-1} \geq 0$ is sufficient for the optimality.
The last condition holds true for $\eta\to 0$ if the light source satisfies
\begin{equation}
 \label{optimal-cond-physical}
 \braket{n} < \frac{g^{(D-1)}(0)}{g^{(D)}(0)}.
\end{equation}
Equations (\ref{optimal-cond-physical-add}) and (\ref{optimal-cond-physical}) form a sufficient condition for the optimality of the $S$-related bounds.
We can obtain the conditions for $S'$-related bounds by replacing $D$ with $D-1$.
The proof of these statements is given in Appendix \ref{appendix-relation-eta0}.
Examples for sources satisfying these conditions are a coherent light source with $\braket{n} < 1$
and a thermal light source with $\braket{n} <  1/D$.

The closed-form expression also allows us to adapt our method easily to the cases where there are ambiguities in the values of $\eta_i$ and $c_{\mathrm{obs},r}$,
simply by calculating the worst-case values and by introducing confidence levels if necessary.
It also helps us to estimate how the degrees of such ambiguities will affect the tightness of the bounds.
As an example, we consider the $D=4$ setup with a uniform efficiency $\eta\ll 1$. 
We assume that the actual distribution $\{p_n\}_n$ is similar to that of a weak coherent light source or a single photon source,
namely, $p_n\gg p_{n+1}$ for $n\geq 1$.
It implies $p_0\sim 1-\tilde{c}_{\mathrm{obs},1}$ and  $p_n\sim \tilde{c}_{\mathrm{obs},n}$ for $n\geq 1$.
Since $c_{r,r}= r!\eta^r$, we have
$\Delta\tilde{c}_{\mathrm{obs},r}/ \tilde{c}_{\mathrm{obs},r} = \Delta c_{\mathrm{obs},r}/ c_{\mathrm{obs},r} - r (\Delta\eta/ \eta).$
By applying it to the dominant terms in Eqs.\;(\ref{eq:explicitD4-pL0})-(\ref{eq:explicitD4-pU3}), we obtain
\begin{equation}
 \label{eq:ambiquity-p0}
 \frac{\Delta p^{U,L}_0}{p_0} \lesssim \frac{p_1}{p_0}\(\frac{|\Delta c_{\mathrm{obs},1}|}{c_{\mathrm{obs},1}}+\frac{|\Delta \eta|}{\eta}\),
\end{equation}
\begin{equation}
 \label{eq:ambiquity-pn}
 \frac{\Delta p^{U,L}_n}{p_n} \lesssim \frac{|\Delta c_{\mathrm{obs},n}|}{c_{\mathrm{obs},n}}+n\frac{|\Delta \eta|}{\eta} \qquad (n\geq 1).
\end{equation}
These show that the relative errors in $c_{\mathrm{obs},r}$ and $\eta$ affect the bounds only proportionally, unless $p_0=0$.
While the above relations are derived from the explicit form for $D=4$, it is not difficult to show that they hold for arbitrary values of $D$.

\section{The decoy-state BB84 protocol with a calibrated source}

As an application of the proposed characterization method,
we consider the calibration of the laser light source used in the decoy-state BB84 protocol,
which is one of the most frequently demonstrated QKD protocols.
Here we focus on the protocol using pulses with three different intensities, termed signal, decoy, and vacuum.
Let $\{p_n\}_n$ and $\{p'_n\}_n$ be the photon-number distributions of the signal and the decoy pulses, respectively.
We assume that the vacuum pulses contain no photons.
The crux of the decoy-state protocol is to estimate the amount of valid signals,
namely, the conditional detection probability $Y_1$ given the sender emits exactly
one photon ($Y_1$ is often called the one-photon yield).
Comparison between the observed probability of the detected signal pulses, $Q$,
and that of the detected decoy pulses, $Q'$, establishes a lower bound $Y_1^L$ on $Y_1$.
Combined with an upper bound $e_1^L$ on the error probability in the single photon emission events,
an asymptotic secure key rate is given by \cite{Gottesman2004}
\begin{align} 
R=&(qp^L_1+q'p'^L_1)Y^L_1(1-H(e^U_1))\nonumber \\ 
&-(Q+Q')H((QE+Q'E')/(Q+Q')). \label{eq:keyrate}
\end{align}
where $q$ is the probability of choosing the signal pulse and $E$ is the observed error probability for the signal pulses,
with $q'$ and $E'$ similarly defined for the decoy pulses.

The security of the protocol has usually been proved under the {\it a priori} Poissonian assumption,
namely, $p_n=\exp(-\mu)\frac{\mu^n}{n!}$ and $p'_n=\exp(-\mu')\frac{\mu'^n}{n!}$ where $\mu$ and $\mu'$
are the mean photon numbers of the signal and the decoy pulses, respectively.
For $1>\mu>\mu'$, the bounds \cite{Wang2009} are then given by 
\begin{equation}
 Y_1^L=\frac{p_2Q'/q'-p'_2Q/q-(p'_0p_2-p_0p'_2)Y_0}
	{p'_1p_2-p_1p'_2} \label{eq:y1_poisson}
\end{equation}
\begin{align}
e_1^U&=(Q'E'/q'-p'_0e_0Y_0)/(p'_1Y_1^L), \label{eq:e1_poisson}
\end{align}
where the zero-photon yield $Y_0$ is determined from the observed
probability of the detected vacuum pulses, and $e_0=1/2$.

Since the Poissonian assumption involves infinite number of conditions, it is impossible to verify it experimentally.
The proposals \cite{Horikiri2006,Wang2007,Adachi2007} for the use of light sources other than lasers also rely on an infinite set of conditions.
It is thus important to replace such {\it a priori} assumptions with experimentally verifiable ones \cite{Zhao2008,Lucamarini2015a}.
What we seek here is to use experimentally available bounds from Theorem 1.
Wang et al. \cite{Wang2009} have shown that Eqs.\;(\ref{eq:y1_poisson}) and (\ref{eq:e1_poisson}) are still 
valid by replacing each term by the worst-case values, provided that 
$ p^L_{n}/p'^U_{n} \geq p^L_{2}/p'^U_{2} \geq p^L_{1}/p'^U_{1} $ holds for all $n\geq 2$.
We can easily extend it to a bound valid when
$ p^L_{n}/p'^U_{n} \geq p^L_{2}/p'^U_{2} \geq p^L_{1}/p'^U_{1} $ holds only for $2\leq n \leq D-1$ as
\begin{align}
 Y_1^L= &\frac{p^L_2Q'/q'-p'^U_2Q/q-(p'^U_0p^L_2-p^L_0p'^U_2)Y_0}
	{p'^U_1p^L_2-p^L_1p'^U_2}\nonumber\\
&-\frac{p^L_2p'^U_{\geq D}}{p'^U_1p^L_2-p^L_1p'^U_2}.
 \label{eq:y1-2} 
\end{align}
Eq.\;(\ref{eq:e1_poisson}) is straightforwardly modified to
 \begin{align}
	\label{eq:e1-2}
e_1^U&=(Q'E'/q'-p'^L_0e_0Y_0)/(p'^L_1Y_1^L).
 \end{align}
The detailed derivation is given in Appendix \ref{appendix-decoy-proof}.

\begin{figure}[h]
	\begin{center}
		\includegraphics[width=7.2cm]{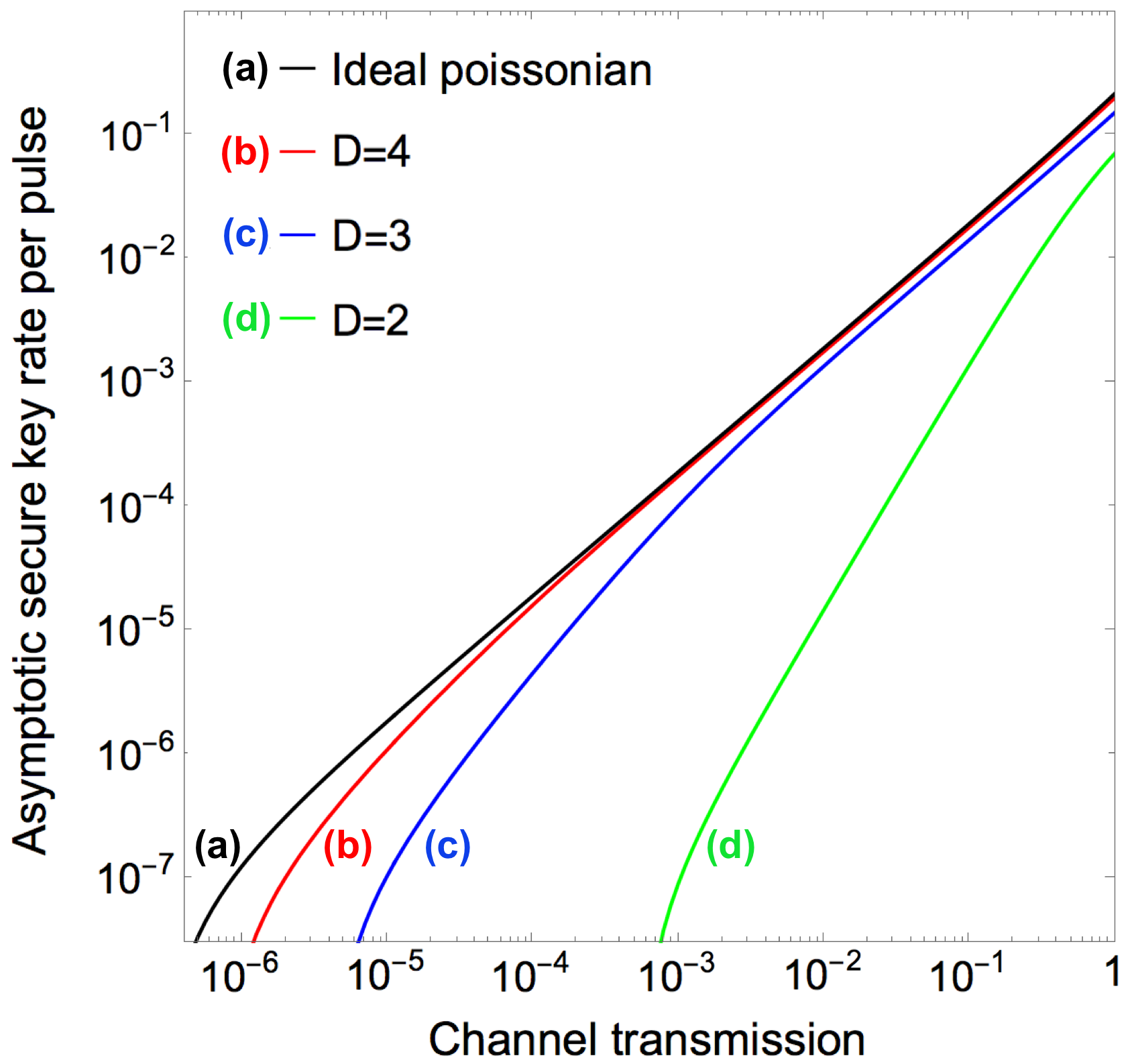}
	 \caption{
	 Asymptotic secure key rates per pulse for the decoy-state BB84 protocol with various assumptions on the light source.
	 (a) An ideal Poissonian source with known distributions $p_{n}=e^{-\mu}\mu^n/n!$ and $p'_{n}=e^{-\mu'}\mu'^n/n!$.
	 (b)-(d) Based on the calibration method with $D=4,3,2$ detectors applied to the same source.
	 For the calibration, we assume that all the overall detection efficiencies $\{\eta_i\}_i$ have the same value $\eta$,
	 and there is $\pm 1\%$	 ambiguity in the value of $\eta$ as $\eta=(0.1 \pm 0.001)/D$.
	 We assume that statistical errors in estimating $c_{\mathrm{obs},r}$ are negligible.
	 For the protocol, we assume $q=0.8$, $q'=0.1$, $Y_0=10^{-8}$ and that the channel causes a constant error of 1\% regardless of the transmission.
	 The detection rate $Q$ and the bit error rate $E$ are modeled as $Q=1-\exp(-\mu\tau)+Y_0$ and $QE =  0.01(1-\exp(-\mu\tau))+ 0.5 Y_0$,
	 where $\tau$ is the channel transmission.
	 $Q'$ and $E'$ are defined similarly.
	 The mean photon numbers are optimized for each value of $\tau$ under the condition $\mu'=\mu/10$.
	 }
		\label{fig:keyrate}
	\end{center}
\end{figure}

Figure \ref{fig:keyrate} shows a comparison of the asymptotic secure key rate with the {\it a priori} assumption of ideal Poissonian and those with our calibration method.
To calculate these curves, we used Eqs.\;(\ref{eq:y1_poisson}) and (\ref{eq:e1_poisson}) for curve (a) and Eqs.\;(\ref{eq:y1-2}) and (\ref{eq:e1-2}) for curves (b) and (c) in Fig.\;\ref{fig:keyrate}.
For curve (d) with $D=2$, Eq.\;(\ref{eq:y1-2}) does not hold, and we used a trivial bound shown in Appendix \ref{appendix-decoy-proof}.
We assume that the overall quantum efficiencies of the detectors are uniform and known with an accuracy of $1$ percent.
From Fig.\;\ref{fig:keyrate}, we find that the secure key rate improves as $D$ increases,
and that for $D=4$ it is comparable to that with the {\it a priori} assumption of ideal Poissonian.

 \section{Summary and discussion}
 We have presented a calibration method for photon-number distribution of a light source and shown the explicit formula for rigorous bounds on probabilities for small photon numbers.
 We believe that our calibration method makes a significant contribution to the quantum optics toolbox,
 and is especially useful in quantum information,
 such as in applications involving security and in computational tasks whose outcome cannot be verified efficiently.

 For the measurement of the photon-number distribution of a light source, other methods are known such as the one using homodyne tomography \cite{Lvovsky2001} and the one using a photon-number-resolving detector \cite{Waks2004,Rosenberg2005,Fujiwara2007}.
 Compared to a threshold detector, these devices can singly respond to two or more photons. 
 This leads to a more compact setup for the calibration, and will be useful for rough estimates.
 On the other hand, these devices tend to require many system parameters to model them, which may make it difficult to produce a rigorous bound.
 They cannot avoid saturation behavior either, namely, a photon-number-resolving detector can resolve photons only up to a certain photon number, and homodyne tomography usually introduces a threshold manually to avoid artifacts.
 On these grounds, we believe that the HBT setup has an advantage of being made up of components, each of which is represented by a simple model.
 Of course, this inevitably raises a question of how closely an actual detector behaves to the threshold detector model.
 An obvious deviation is the dark counting, which may be treated as an estimation problem of genuine coincidence rates from the actually observed coincidence rates including the contribution of the dark counts.
 It is also important to verify the validity of the threshold detector model in future experimental researches.



\begin{acknowledgements}
 This work was funded in part by ImPACT Program of Council for
Science, Technology and Innovation (Cabinet Office, Government of Japan), Photon Frontier Network
 Program (Ministry of Education, Culture, Sports, Science and Technology), and CREST (Japan Science and Technology Agency).
\end{acknowledgements}

\appendix
\section{}
\label{appendix-proof-theorem}
We justify inequalities (\ref{eq:puodd}), (\ref{eq:pleven}) and (\ref{eq:pld}) involving the index set $S'$.
The proof of Eq.\;(\ref{eq:pld}) proceeds exactly as that of Eq.\;(\ref{eq:pud}) in the main text by setting
$z_j= {\bm \omega}_j\cdot {\bm d}^{(S')}_\infty$,
except that $f(D)=1$ is replaced by $\lim_{m\to \infty} f(m)=1$.
Combined with $f(D-1)=0$, we have $f'(m)>0$ for $m\geq D-1$
and hence $f(m)<1$ for $m\geq D$. This leads to Eq.\;(\ref{eq:pld}).

The proof of Eqs.\;(\ref{eq:puodd}) and (\ref{eq:pleven}) proceeds as follows.
By setting $z_j =\bm{\omega}_j \cdot\bm{d}^{(S')}_{n}$ in Eq.\;$(\ref{eq:fm})$ for a fixed value of $n\in \{0,\ldots, D-1\}$, we see that $f(m)=\bm{c}_{m}\cdot\bm{d}^{(S')}_{n}$ for nonnegative integer $m$. Since ${\bm \omega}_0=(1,1,\ldots,1)=\bm{c}_{\infty}$, we have $z_0= \bm{c}_{\infty} \cdot\bm{d}^{(S')}_{n}=0$. If $\{z_j\}_{j=1,\ldots,D}$ are nonzero and have the same sign, it follows that $ {D}_{\beta_{D-1}}\cdots {D}_{\beta_{1}} f(m)$ has no zeros and $f(m)$ has no more than $D-1$ zeros. If some of $\{z_j\}_{j=1,\ldots,D}$ are zero or have the opposite sign, $f(m)$ has fewer zeros. Since $f(m)=0$ for $m\in \{0,\ldots,D-1\} /\{n\}$, the number of zeros of $f(m)$ is $D-1$. It also means that the number of zeros of ${D}_{\beta_{1}} f(m)$ is $D-2$, and ${D}_{\beta_{1}} f(m) \neq 0$ at zeros of $f(m)$. This property and $f'(m)={D}_{\beta_{1}} f(m)-{\beta_{1}} f(m)$ implies that $f'(m)$ is non-zero at zeros of $f(m)$. When $D-1-n$ is even, $f(m)$ ($m>D-1$) is positive, leading to Eq.\;(\ref{eq:puodd}).  When $D-1-n$ is odd, $f(m)$ ($m>D-1$) is negative, leading to Eq.\;(\ref{eq:pleven}).

 \section{}
 \label{appendix-explicit-formula}
In this Appendix, we present explicit formulas for the bounds calculated from Theorem 1 in the case of $D=2, 3,$ and $4$.
To simplify the notations, we define $\tilde{c}_{{\rm obs},r}:=c_{{\rm obs},r}/c_{r,r}$, $s_j:=\sum_{W\in I_j}\prod_{i\in W}\eta_i/\binom{D}{j}$ ($j=2,\ldots,D$), and $\xi_{i,j}:=s_i/(s_{j}\eta^{i-j})-1$ ($i,j=2,\ldots,D$). The formula for the uniform case of $\eta=\eta_1=\eta_2=\eta_3=\eta_4$ is simply given
by setting $\xi_{i,j}=0$ for all $i,j$.\\
\\
($D=2$)
\begin{align}
   p^L_{0}=&1-\tilde{c}_{{\rm obs},1}+2(1+\xi_{2,1})\eta(1-\eta)\tilde{c}_{{\rm obs},2}  , \\
   p^U_{0}=&1-\tilde{c}_{{\rm obs},1}+[1-(1 - \xi_{2,1})\eta]\tilde{c}_{{\rm obs},2}   , \\
   p^L_{1}=&\tilde{c}_{{\rm obs},1}-[2-(1 - \xi_{2,1})\eta]\tilde{c}_{{\rm obs},2} , \\
 p^U_{1}=&\tilde{c}_{{\rm obs},1}-2(1+\xi_{2,1})\eta\tilde{c}_{{\rm obs},2},\\
 p^L_{\geq 2}=&2!(1+\xi_{2,1})\eta^2\tilde{c}_{{\rm obs},2},\\
 p^U_{\geq 2}=&\tilde{c}_{{\rm obs},2}.
\end{align}
($D=3$)
\begin{align}
   p^L_{0}=&1-\tilde{c}_{{\rm obs},1}+[1 -  (1 - 2 \xi_{2,1})\eta]\tilde{c}_{{\rm obs},2}\nonumber \\
   &-[1 - (3 - 3 \xi_{3,2}/2)  \eta+ (2 - 9 \xi_{3,2}/2 \nonumber \\
   & + 2 \xi_{3,1})\eta^2] \tilde{c}_{{\rm obs},3}  , \\
   p^U_{0}=&1-\tilde{c}_{{\rm obs},1}+[1 -  (1 - 2 \xi_{2,1})\eta]\tilde{c}_{{\rm obs},2}\nonumber \\
   &-3 (1+\xi_{3,2})\eta (1 -3 \eta+2(1+ \xi_{2,1}) \eta^2 )\tilde{c}_{{\rm obs},3}   , \\
   p^L_{1}=&\tilde{c}_{{\rm obs},1}-[2- (1 - 2 \xi_{2,1})\eta]\tilde{c}_{{\rm obs},2}\nonumber \\
   &+3(1+\xi_{3,2})\eta(2-3\eta)\tilde{c}_{{\rm obs},3}  , \\
   p^U_{1}=&\tilde{c}_{{\rm obs},1}-(2-\eta)\tilde{c}_{{\rm obs},2}+[3-(6- 3 \xi_{3,2})\eta\nonumber \\
   &+(2- 9/2 \xi_{3,2} + 2 \xi_{3,1})\eta^2]\tilde{c}_{{\rm obs},3}   , \\
   p^L_{2}=&\tilde{c}_{{\rm obs},2}-3[1-(1 - \xi_{3,2}/2)\eta]\tilde{c}_{{\rm obs},3}  ,\\
 p^U_{2}=&\tilde{c}_{{\rm obs},2}-3(1+\xi_{3,2})\eta\tilde{c}_{{\rm obs},3},\\
 p^L_{\geq 3}=&3!(1+\xi_{3,1})\eta^3\tilde{c}_{{\rm obs},3},\\
 p^U_{\geq 3}=&\tilde{c}_{{\rm obs},3}.
\end{align}
\\
($D=4$)
\begin{align}
   p^L_{0}=&1-\tilde{c}_{{\rm obs},1}+[1- (1 - 3 \xi_{2,1}) \eta]\tilde{c}_{{\rm obs},2}\nonumber \\
   &-[1 -  (3 - 3 \xi_{3,2})\eta +  (2 - 12 \xi_{3,2} + 6 \xi_{3,1})\eta^2]\tilde{c}_{{\rm obs},3}\nonumber \\
   &+4 (1+\xi_{4,3}) \eta [1 - 6 \eta +  (11 + 3 \xi_{3,1})\eta^2 \nonumber \\
   &- 6  (1+\xi_{3,1}) \eta^3]
   \tilde{c}_{{\rm obs},4}  , \label{eq:explicitD4-pL0} \\
   p^U_{0}=&1-\tilde{c}_{{\rm obs},1}+[1- (1 - 3 \xi_{2,1}) \eta]\tilde{c}_{{\rm obs},2}\nonumber \\
   &-[1 -  (3 - 3 \xi_{3,2})\eta +  (2 - 12 \xi_{3,2} + 6 \xi_{3,1})\eta^2]\tilde{c}_{{\rm obs},3}\nonumber \\
   &+[1-(6 - 2 \xi_{4,3})\eta+(11 + 6 \xi_{2,1} - 8 \xi_{3,2}/3 - 12 \xi_{4,3} \nonumber \\
   &+11 \xi_{4,2}/3)\eta^2-(5+24 \xi_{2,1} - 32 \xi_{3,2}/3 - 16 \xi_{4,3} \nonumber \\
   &+44 \xi_{4,2}/3 - 6 \xi_{4,1})\eta^3]\tilde{c}_{{\rm obs},4}, \\
   p^L_{1}=&\tilde{c}_{{\rm obs},1}-[2 - (1 - 3 \xi_{2,1})\eta]\tilde{c}_{{\rm obs},2} \nonumber \\
   &+[3 - (6 - 6 \xi_{3,2}) \eta +  (2 - 12 \xi_{3,2} + 6 \xi_{3,1})\eta^2]\tilde{c}_{{\rm obs},3}\nonumber \\
   &-[4 - (18 - 6 \xi_{4,3}) \eta + (22 + 12 \xi_{2,1} - 16 \xi_{3,2}/3 \nonumber \\
   &- 24 \xi_{4,3} +  22 \xi_{4,2}/3) \eta^2 - (6+24 \xi_{2,1} - 32 \xi_{3,2}/3  \nonumber \\
   &- 16 \xi_{4,3} +  44 \xi_{4,2}/3 - 6 \xi_{4,1}) \eta^3]\tilde{c}_{{\rm obs},4}  , \\
   p^U_{1}=&\tilde{c}_{{\rm obs},1}-[2 - (1 - 3 \xi_{2,1})\eta]\tilde{c}_{{\rm obs},2} \nonumber \\
   &+[3 - (6 - 6 \xi_{3,2}) \eta +  (2 - 12 \xi_{3,2} + 6 \xi_{3,1})\eta^2]\tilde{c}_{{\rm obs},3}\nonumber \\
   &- 4(1+\xi_{4,3})\eta[3  - 12 \eta+ (11 + 3 \xi_{3,1})\eta^2 ]\tilde{c}_{{\rm obs},4}   , \\
   p^L_{2}=&\tilde{c}_{{\rm obs},2}-3[1 - (1 - \xi_{3,2}) \eta]\tilde{c}_{{\rm obs},3}\nonumber \\
   &+12(1+ \xi_{4,3})\eta(1  - 2 \eta)\tilde{c}_{{\rm obs},4}  ,\\
   p^U_{2}=&\tilde{c}_{{\rm obs},2}-3[1 - (1 - \xi_{3,2}) \eta]\tilde{c}_{{\rm obs},3}\nonumber \\
   &+[6-(18 - 6 \xi_{4,3})\eta+(11 + 6 \xi_{2,1} - 8 \xi_{3,2}/3 \nonumber \\
   &- 12 \xi_{4,3} + 11 \xi_{4,2}/3)\eta^2]\tilde{c}_{{\rm obs},4}, \\
   p^L_{3}=&\tilde{c}_{{\rm obs},3}-[4 - 2  (3 - \xi_{4,3})\eta]\tilde{c}_{{\rm obs},4}  ,\\
 p^U_{3}=&\tilde{c}_{{\rm obs},3}-4(1+\xi_{4,3})\eta\tilde{c}_{{\rm obs},4},\label{eq:explicitD4-pU3}\\
 p^L_{\geq 4}=&4!(1+\xi_{4,1})\eta^4\tilde{c}_{{\rm obs},4},\label{eq:explicitD4-pL4}\\
 p^U_{\geq 4}=&\tilde{c}_{{\rm obs},4}.\label{eq:explicitD4-pU4}
\end{align}

\section{}
\label{appendix-relation-eta0}
In this Appendix, we investigate properties of the bounds appearing in Theorem 1 in the limit of $\eta\to 0$.
We assume that the efficiencies are uniform ($\eta_i=\eta$).
Then it is not difficult to show $c_{n,r}/\eta^{-r}$ $\to$ $n!/(n-r)!$ for $\eta\to 0$,
since $c_{n,r}$ is equal to the probability for at least $r$ out of $n$ photons to survive to reach $r$ detectors.
We also exclude sources which are singular. More precisely,
we assume that the factorial moment of order $D$,
\begin{equation}
 \braket{(n)_D} := \sum_{n} n(n-1)\cdots(n-D+1)p_n
\end{equation}
is finite and $c_{\text{obs}, r}=O(\eta^r)$.


We first discuss the relation between the $S'$-related bounds for a $D$-detector setup and the $S$-related bounds for a $(D-1)$-detector setup.
We denote the former as $p_n^{(S',D)}:=\bm{c}_{\text{obs}}\cdot \bm{d}_n^{(S')}$ and the latter as $p_n^{(S,D-1)}:=\bm{c}_{\text{obs}}\cdot \bm{d}_n^{(S)}$.
From the biorthogonality relations, we have
\begin{equation}
 \label{eta0-cobs-def-D}
 \sum_{n=0}^{D-1} c_{n,r} p_n^{(S',D)}  + p_\infty^{(S',D)}=c_{\text{obs}, r}
\end{equation}
\begin{equation}
 \label{eta0-cobs-def-D-1}
 \sum_{n=0}^{D-1} c_{n,r} p_n^{(S,D-1)} = c_{\text{obs}, r}
\end{equation}
for $r=1,\ldots, D-1$.
They also formally hold true for $r=0$ if we define $c_{n,0}:=1$ and $c_{\text{obs},0}:=1$.
From Eq.\;(\ref{eta0-cobs-def-D}) with $r=D$, we have 
$p_\infty^{(S',D)}=c_{\text{obs}, D}$.
By subtracting Eq.\;(\ref{eta0-cobs-def-D}) from Eq.\;(\ref{eta0-cobs-def-D-1}),
we obtain 
\begin{equation}
\sum_{n=0}^{D-1}
\eta^{-r} c_{n,r} (p_n^{(S,D-1)}-p_n^{(S',D)})
= \eta^{-r} c_{\text{obs}, D}  
\end{equation}
for $r=0,\ldots,D-1$. Notice that the right-hand side is at most $O(\eta)$.
Since $\{\eta^{-r} c_{n,r}\}_{n,r}$ converge to a triangular matrix with nonzero diagonal elements for $\eta\to 0$, we see that
\begin{equation}
 \label{eta0-equivalence-S-Sdash}
 p_n^{(S,D-1)}-p_n^{(S',D)} = O(\eta),
\end{equation}
implying that the $S$-related bounds for $(D-1)$-detector setup improves only by $O(\eta)$ through adding another detector.

Next, we derive a sufficient condition for the source to make all the bounds non-negative and hence to make the bounds optimal.
Since $c_{m,r}\eta^{-r}$ converges to $m!/(m-r)!$, which is a polynomial of $m$ of order $r$,
so does $\bm{c}_{m}\cdot \bm{d}^{(S)}_n$ to a polynomial of order at most $D$, which we denote by $f(m)$.
The function $f(m)$ satisfies $f(m)=0$ for $0\leq m \leq D$ except for $m=n$ and $f(n)=1$.
The former condition means that $f(m)\propto m(m-1)\cdots(m-D)/(m-n)$.
The latter condition means that the normalization factor is $(-1)^{D-n}/(n!(D-n)!)$.
We thus obtain
\begin{equation}
 \label{cmdn-eta0}
 \bm{c}_{m}\cdot \bm{d}^{(S)}_n \to  \frac{m(m-1)\cdots(m-D)}{m-n} \frac{(-1)^{D-n}}{n!(D-n)!}.
\end{equation}
Since $\bm{c}_{\text{obs}}\cdot\bm{d}^{(S)}_n=\sum_{m\geq 0}\bm{c}_{m}\cdot\bm{d}^{(S)}_n  p_m$,
we have
\begin{equation}
	\bm{c}_{\text{obs}}\cdot\bm{d}^{(S)}_n
	\to p_n - \sum_{m\geq D+1}\frac{m(m-1)\cdots (m-D)}{m-n}\chi_n p_m
\end{equation}
for $n=D-1, D-3,\ldots,$ where $\chi_n:=1/(n!(D-n)!)$.
Comparison of its right-hand side with the one with $n=D-1$, we see that
\begin{equation}
 \begin{split}
	\bm{c}_{\text{obs}}\cdot\bm{d}^{(S)}_n
	\geq& p_n - \frac{\chi_n}{\chi_{D-1}}(p_{D-1}-\bm{c}_{\text{obs}}\cdot\bm{d}^{(S)}_{D-1})\\
	= & p_n - \frac{1}{D}\binom{D}{n}(p_{D-1}-\bm{c}_{\text{obs}}\cdot\bm{d}^{(S)}_{D-1})	
 \end{split}
\end{equation}
holds for $\eta\to 0$.
It means that if Eq.\;(\ref{optimal-cond-physical-add}) holds, positivity of $\bm{c}_{\text{obs}}\cdot \bm{d}^{(S)}_n$ follows that of $\bm{c}_{\text{obs}}\cdot \bm{d}^{(S)}_{D-1}$.

For $n=D-1$, Eq.\;(\ref{optimal-cond-physical}) can be rewritten as
$\bm{c}_{\text{obs}}\cdot \bm{d}^{(S)}_{D-1} \to (\braket{(n)_{D-1}}-\braket{(n)_{D}})/(D-1)!$.
Since $g^{(r)}(0)$ is $\braket{(n)_r}/\braket{n}^r$, we see that if Eq.\;(\ref{optimal-cond-physical}) holds,
$\bm{c}_{\text{obs}}\cdot \bm{d}^{(S)}_n > 0$ in the limit of $\eta\to 0$.
Due to Eq.\;(\ref{eta0-equivalence-S-Sdash}), the whole argument is applicable to the $S'$-related bounds just by replacing $D$ by $D-1$.

Next, we consider how these conditions are applied to typical light sources.
For a coherent light source, we have $p_n=\exp(-\braket{n})\braket{n}^n/n!$ and $g^{(r)}(0) = 1$,
and Eq.\;(\ref{optimal-cond-physical}) becomes $\braket{n} < 1$.
For a thermal light source with $p_n=\frac{1}{\braket{n}+1}\(\frac{\braket{n}}{\braket{n}+1}\)^n$, we have $g^{(r)}(0)=r!$
and Eq.\;(\ref{optimal-cond-physical}) becomes $\braket{n} < 1/D$.
In these two cases, we can also check that Eq.\;(\ref{optimal-cond-physical-add}) holds if Eq.\;(\ref{optimal-cond-physical}) is satisfied.
On the other hand, the condition (\ref{optimal-cond-physical-add}) is not fulfilled by a pseudo single-photon source with an exceptionally small value of $p_0$.
While the optimality may not be achieved then, practically we will seldom encounter such a source due to coupling losses in experiments.

\section{}
\label{appendix-decoy-proof}
In this Appendix, we justify the lower bound $Y_1^L$ given in Eq.\;(\ref{eq:y1-2}) and the upper bound $e_1^U$ in Eq.\;(\ref{eq:e1-2}).
We assume $D\geq 3$ and that $p^L_n/p'^U_n\geq p^L_2/p'^U_2 \geq p^L_1/p'^U_1$ for $2\leq n \leq D-1$  are verified by the calibration.
The conditions $Q=q\sum_{n}p_nY_n$, $Q'=q'\sum_np'_nY_n$ and $0\leq Y_n \leq 1$ leads
\begin{equation}
 \label{eq:appendixB-Qdash-qdash}
 \frac{Q'}{q'} \leq \sum_{n=0}^{D-1}p'^U_nY_n + p'^U_{\geq D}
\end{equation}
and
\begin{equation}
 \frac{Q}{q} \geq \sum_{n=0}^{D-1}p^L_nY_n.
\end{equation}
Combined with $p^L_n/p'^U_n\geq p^L_2/p'^U_2$ for $2\leq n \leq D-1$, we have
\begin{equation}
 p_2^L \frac{Q'}{q'} - p'^U_2\frac{Q}{q} \leq \sum_{n=0,1} \(p^L_2p'^U_n-p'^U_2p^L_n\)Y_n + p_2p'^U_{\geq D}.
\end{equation}
Since $p^L_2/p'^U_2 \geq p^L_1/p'^U_1$, we obtain Eq.\;(\ref{eq:y1-2}).

For $D=2$, we obtain a bound $Y^L_1$ directly from Eq.\;(\ref{eq:appendixB-Qdash-qdash}) as
\begin{equation}
 Y_1^L = \(\frac{Q'}{q'}-p'^U_0Y_0-p'^U_{\geq D}\)/p'^U_1.
\end{equation}

For the derivation of $e^U_1$, we use $Q'E'=q'\sum_{n} p'_nY_ne_n$ and $Y_n\geq 0$, leading to
\begin{equation}
 \frac{Q'E'}{q'} \geq p'^L_0Y_0e_0 + p'^L_1Y_1e_1,
\end{equation}
and obtain Eq.\;(\ref{eq:e1-2}).

\end{document}